\documentclass[
	a4paper, 
	10pt, 
	unnumberedsections, 
	twoside, 
]{LTJournalArticle}

\runninghead{FrackyFrac: A Standalone UniFrac Calculator} 

\footertext{} 

\title{FrackyFrac: A Standalone UniFrac Calculator}

\author{%
	Amit Lavon\textsuperscript{1,2}, Smadar Shilo\textsuperscript{1,2,3,4}, Ayya Keshet\textsuperscript{1,2} and Eran Segal\textsuperscript{1,2}\thanks{Corresponding author: \href{mailto:eran.segal@weizmann.ac.il}{eran.segal@weizmann.ac.il}}
}

\date{\footnotesize
\textsuperscript{\textbf{1}}Department of Computer Science and Applied Mathematics, Weizmann Institute of Science, Rehovot, Israel\\
\textsuperscript{\textbf{2}}Department of Molecular Cell Biology, Weizmann Institute of Science, Rehovot, Israel\\
\textsuperscript{\textbf{3}}The Jesse Z and Sara Lea Shafer Institute for Endocrinology and Diabetes, National Center for Childhood Diabetes, Schneider Children’s Medical Center of Israel, Petah Tikva, Israel\\
\textsuperscript{\textbf{4}}Faculty of Medicine, Tel Aviv University, Tel-Aviv, Israel}

\renewcommand{\maketitlehookd}{%
	\begin{abstract}
		\noindent UniFrac is a family of distance metrics over microbial abundances, that take taxonomic relatedness into account. Current tools and libraries for calculating UniFrac have specific requirements regarding the user’s technical expertise, operating system, and pre-installed software, which might exclude potential users. FrackyFrac is a native command-line tool that can run on any platform and has no requirements. It can also generate the phylogenetic trees required for the calculation. We show that FrackyFrac’s performance is on par with currently existing implementations. FrackyFrac can make UniFrac accessible to researchers who may otherwise skip it due to the effort involved, and it can simplify analysis pipelines for those who already use it.
	\end{abstract}
}

\begin{document}

\maketitle

\section{Introduction}

The study of microbial communities involves analyzing their biodiversity. The two axes of diversity typically examined are alpha diversity and beta diversity. Alpha diversity measures the biodiversity within a single community, while beta diversity measures the diversity between different communities. Measures of beta diversity allow the comparison of microbial communities and facilitate downstream analysis such as clustering and visualization.

The Bray-Curtis dissimilarity index \cite{bray_ordination_1957} is commonly used to measure beta diversity by comparing microbial abundances. This index does not use external information about the compared operational taxonomic units (OTUs). The formula is simple and easy to calculate, and thus it is included in many studies.

UniFrac \cite{lozupone_unifrac_2005} is a distance metric over microbial abundance vectors that takes into account the phylogenetic relatedness between OTUs. It uses a user-provided phylogenetic tree and returns the Jaccard similarity (or dissimilarity) of the samples over the branches of the tree. This way, two different OTUs are in agreement over the part of the tree that they share, starting from the root and until they diverge to separate branches, and in disagreement over the parts of the tree past their divergence. The closer the OTUs are on the tree, the higher their similarity. Two flavors of UniFrac that are commonly used are weighted and unweighted. In weighted UniFrac, every species contributes to the total similarity relative to its abundance, while in unweighted UniFrac every species has the same weight.

The advantage of UniFrac is in the differential distances between the OTUs, according to their taxonomic relation. Similar environments in different geographies may harbor similar makeups of microbial families, orders, classes, etc. even if the individual species are different. These similarities are lost when using Bray-Curtis, but are kept when using UniFrac, allowing an informative comparison of environments.
Over time, improvements and variations to the UniFrac metric were developed. Generalized UniFrac \cite{chen_associating_2012}, Variance-Adjusted Weighted UniFrac \cite{chang_variance_2011}, Information UniFrac and Ratio UniFrac \cite{wong_expanding_2016} were created to broaden the kinds of variance that can be measured. Performance improvements such as Fast UniFrac \cite{hamady_fast_2010} were introduced to allow UniFrac to run on larger datasets. Recent advancements have introduced the Striped UniFrac \cite{mcdonald_striped_2018} algorithm and GPU support \cite{sfiligoi_optimizing_2022}, which increased the speed of computation to make it applicable to datasets with over 100K samples.

These modern implementations come with specific challenges. R and Python-based implementations require having these languages installed, installing third-party libraries, and having some level of proficiency with these languages. Some of those third-party libraries, such as OpenMP and HDF5, require setting up the environment in which the program runs in a specific way. Bioconda’s implementation works with the BIOM and QZA formats, which are specific to that framework. Some of the implementations do not support Windows. Each such constraint adds friction on the way to getting UniFrac in the analysis pipeline. Scientists may give up on using UniFrac in their research if it requires more energy or expertise than what they have.

To demonstrate the growth potential for UniFrac, we searched Google Scholar for papers that mention Bray-Curtis without mentioning UniFrac, published in 2022 with source “nature”\footnote{Query: "bray curtis" -unifrac source:nature} (which included journals from the Nature Publishing Group) and got 562 results. Searching for papers that mention UniFrac, published in 2022 with source “nature”\footnote{Query: unifrac source:nature} yielded 282 results. Under the assumption that UniFrac is relevant to most of the studies mentioning Bray-Curtis, this suggests a potential to double the use of UniFrac.

Reviewing the latest 20 results from the latter search (papers that mention UniFrac), we found that UniFrac was used in studies with sample sizes ranging from 5 to 3211 samples. In 19 out of the 20 papers, sample sizes ranged from 5 to 247 samples, with an average of 69±66. Interestingly, 19 out of 20 of these papers used only the original variations of UniFrac - weighted and unweighted (supplementary table \ref{tab:papers}).

We infer that an inclusive implementation of UniFrac, one that is easily accessible, can substantially expand the usage of the UniFrac metric. From the surveyed studies we learn that supporting the weighted and unweighted variations, with reasonable performance on up to a few thousands of samples, can satisfy most currently published studies.

Here, we set out to create a more accessible implementation of the UniFrac metric. We created a toolkit that is optimized for ease of access and ease of use, and can be used on any platform without installation.

\section{Results}

\subsection{A standalone implementation}

We evaluated our implementation against criteria that may affect a method’s usability (table \ref{tab:deps}). We refer to installation processes as “builtin” and “third party”, where a builtin installation is one that uses a language’s builtin package installation pipeline. Third-party installation mechanisms introduce more moving parts to the process, and are typically tested and maintained by smaller communities. This can lead to more fragile installation processes which take more debugging and therefore might exclude users.

\begin{table*}[h]
    \small
    \centering
    \begin{tabular}{|L{0.13\linewidth}|L{0.17\linewidth}|L{0.2\linewidth}|L{0.2\linewidth}|L{0.2\linewidth}|}
    \hline
    \textbf{Implementation} & \textbf{Supported operating systems} & \textbf{Required programming language} & \textbf{Required pre-existing software} & \textbf{Installation} \\
    \hline
    FrackyFrac & All & None & None & None \\
    \hline
    Phyloseq \cite{mcmurdie_phyloseq_2012} & All & R & R libraries & Third party, with BioConductor \\
    \hline
    QIIME2 \cite{bolyen_reproducible_2019} & Linux, macOS & Anaconda Python & >400 Python libraries & Builtin, with a conda configuration file \\
    \hline
    Scikit-bio & Linux, macOS & Python & Python libraries & Builtin, with pip \\
    \hline
    Bioconda-binaries (SSU) \cite{mcdonald_striped_2018} & Linux, macOS & Anaconda Python & Python libraries & Builtin, with conda \\
    \hline
    \end{tabular}
    \caption{Comparison of UniFrac implementations.}
    \label{tab:deps}
\end{table*}

\subsection{Performance}

\begin{figure}[h]
    \centering
    \includegraphics[width=1\linewidth]{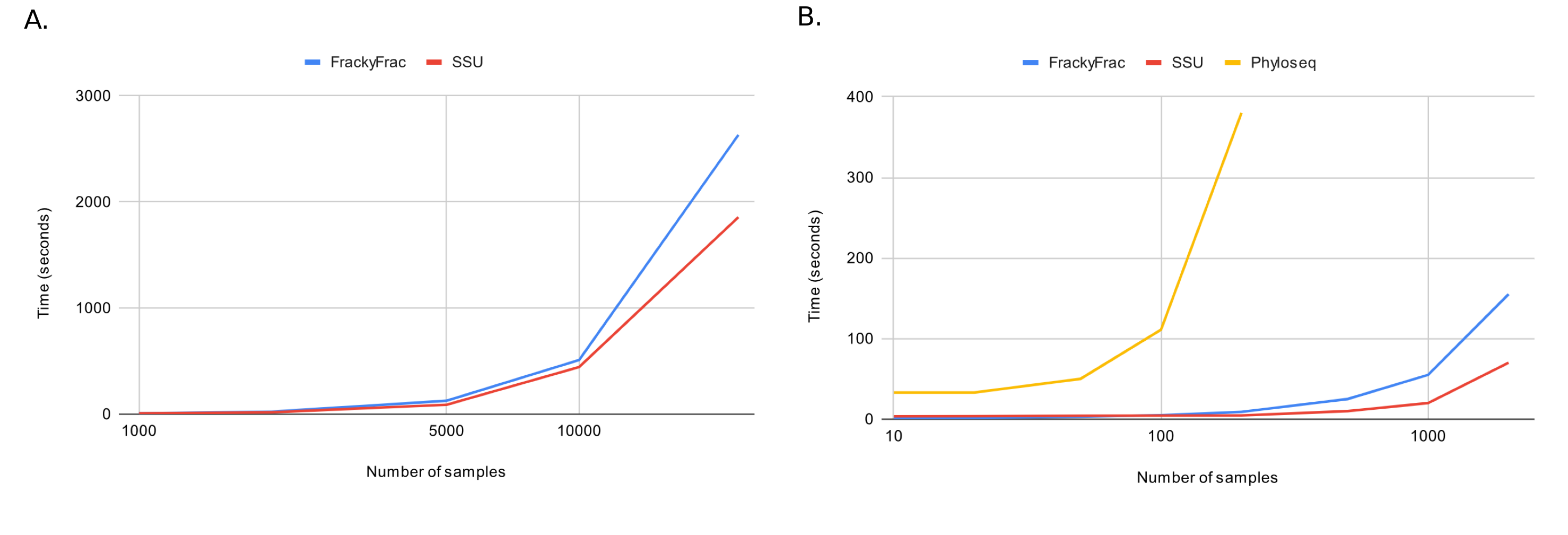}
    \caption{\textbf{UniFrac computation times on Earth Microbiome Project data.} (A) Computation times on 10, 20, 50, 100, 200, 500, 1000 and 2000 samples using a single core. (B) Computation times on 1000, 2000, 5000, 10000 and 23000 (all) samples using 8 cores.}
    \label{fig:perf}
\end{figure}

We ran FrackyFrac on the OTU tables of samples from the Earth Microbiome Project \cite{gilbert_earth_2014} (the “silva” reference was used). To compare run times on different input sizes, subsamples of the data were taken arbitrarily. We compared FrackyFrac with SSU which is the implementation used in QIIME2, and with Phyloseq. Scikit-bio was left out because QIIME2 is a faster Python library. FrackyFrac’s computation times were higher than SSU’s by a factor of about two on a single core, and by a factor of 1.4 on eight cores. Figure \ref{fig:perf} shows the performance of the compared implementations.

\subsection{Validation}

\begin{figure}[h]
    \centering
    \includegraphics[width=1\linewidth]{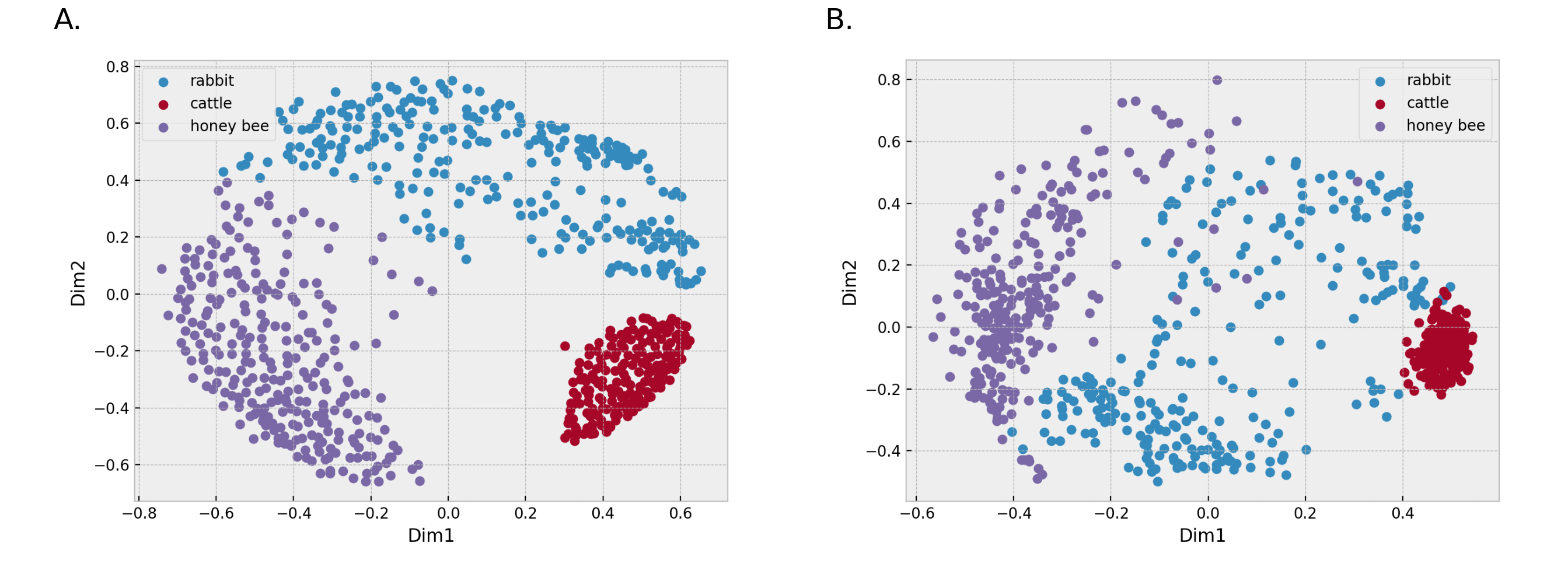}
    \caption{\textbf{PCoA of samples from different host species.} Distances were calculated FrackyFrac’s unweighted (A) and weighted (B) UniFrac, on Earth Microbiome Project samples.}
    \label{fig:scat}
\end{figure}

FrackyFrac’s result distances from the previous section were compared with SSU’s and were found to be equal. The published repository contains a directory named ‘testdata’ with runnable tests and comparison scripts. Figure \ref{fig:scat} shows the separation of samples from several hosts from the Earth Microbiome Project, using FrackyFrac’s weighted and unweighted UniFrac.

\subsection{Creating phylogenetic trees}

FrackyFrac features a builtin tree maker that uses the method suggested by \cite{ondov_mash_2016}, Mash distance with UPGMA clustering, as a simple way to create phylogenetic trees. Since UniFrac requires a phylogenetic tree for its calculation, we provide this method in case no such tree is available.

\section{Conclusions}

In this work we introduce a standalone implementation of weighted and unweighted UniFrac. This evolved from a need to use UniFrac without having to go through software installation, format conversions, or having to switch programming languages or operating systems. FrackyFrac meets that need. Our implementation addresses mainly scientists who are not already versed in one of the existing frameworks for running UniFrac, but also scientists who have been using UniFrac and would like more freedom as to how and where they can use it. Faster alternatives exist, but for the sample sizes common in current studies, calculation speed has less impact than the human factor.

Some aspects are not covered by this work. In terms of scalability, it does not run on GPUs, making it less suitable for very large datasets. In terms of variation, it supports only weighted and unweighted UniFrac. As shown in the introduction, papers citing UniFrac tend to use the two original variations, so supporting only them covers the vast majority of use cases. Future improvements may include supporting additional flavors of UniFrac, and adding optional integration with third-party libraries for better scalability.

FrackyFrac has the potential to increase the use of UniFrac by reducing the amount of effort and expertise required to use it. It can expand the range of potential users to those with more diverse backgrounds and to more platforms.

\section{Implementation}

The FrackyFrac suite contains two executables: frcfrc which takes as input a TSV of abundances and a phylogenetic tree, and outputs pairwise UniFrac distances; and trtr which creates phylogenetic trees given input fasta sequences.

To support different platforms, FrackyFrac is cross-compiled into standalone binaries which contain all the required logic, with a native binary for every operating system. This removes the need for an installation in favor of downloading ready-to-use executables. There is no need to get FrackyFrac’s code in order to use it, so no particular programming language is required on the user’s side.

To support diverse pipelines, FrackyFrac uses the tab-separated-values (TSV) format for input and output, for which there exist encoders and decoders in most modern programming languages.

The algorithm for calculating UniFrac distances is similar to Fast UniFrac’s algorithm, where samples are represented as lists of tree-nodes and abundances. FrackyFrac uses the biostuff package suite \cite{lavon_fluhusbiostuff_2022} for processing biological data and the gostuff package suite \cite{lavon_fluhusgostuff_2022} for algorithms and data structures.

\section{Availability and requirements}

Software: https://github.com/fluhus/frackyfrac. Operating systems: Linux, Windows, MacOS. Software requirements: none. License: MIT. Non-academic use is unrestricted.

\section{Declarations}

\subsection{Availability of data and materials}
The Earth Microbiome Project data that was used for demonstration can be downloaded from:\\
https://earthmicrobiome.org/data-and-code/

\subsection{Competing interests}
The authors declare that they have no competing interests.

\subsection{Authors' contributions}
AL: conceptualization, methodology, software, writing - original draft.
SS, AK: validation, data curation, writing - review \& editing.
ES: supervision, writing - review \& editing.

\printbibliography

\newpage

\section{Supplementary material}

\beginsupplement

\begin{table}[h!]
    \small
    \centering
    \begin{tabular}{|L{0.15\linewidth}|L{0.15\linewidth}|L{0.6\linewidth}|}
    \hline
    Sample size & Type of UniFrac used & Paper title \\
    \hline
    12 & W & White spot syndrome virus impact on the expression of immune genes and gut microbiome of black tiger shrimp Penaeus monodon \cite{jatuyosporn_white_2023} \\
    \hline
    48 & U & Obstructive sleep apnea is related to alterations in fecal microbiome and impaired intestinal barrier function \cite{li_obstructive_2023} \\
    \hline
    64 & U,W,Generalized & Diagnostic and prognostic potential of the microbiome in ovarian cancer treatment response \cite{asangba_diagnostic_2023} \\
    \hline
    160 & W & Exacerbation of allergic rhinitis by the commensal bacterium Streptococcus salivarius \cite{miao_exacerbation_2023} \\
    \hline
    130 & U,W & Altered gut microbiota in individuals with episodic and chronic migraine \cite{yong_altered_2023} \\
    \hline
    20 & U & Rectal swabs as a viable alternative to faecal sampling for the analysis of gut microbiota functionality and composition \cite{radhakrishnan_rectal_2023} \\
    \hline
    64 & U,W & High-fat diet and estrogen modulate the gut microbiota in a sex-dependent manner in mice \cite{hases_high-fat_2023} \\
    \hline
    84 & Not specified & Interactions between perceived stress and microbial-host immune components: two demographically and geographically distinct pregnancy cohorts \cite{penalver_bernabe_interactions_2023} \\
    \hline
    19 & U,W & Severe, short-term sleep restriction reduces gut microbiota community richness but does not alter intestinal permeability in healthy young men \cite{karl_severe_2023} \\
    \hline
    21 & W & Stable colonization of Akkermansia muciniphila educates host intestinal microecology and immunity to battle against inflammatory intestinal diseases \cite{wang_stable_2023} \\
    \hline
    10 & U,W & Long-term taxonomic and functional stability of the gut microbiome from human fecal samples \cite{kim_long-term_2023} \\
    \hline
    5 & Not specified & The rhizospheric bacterial diversity of Fritillaria taipaiensis under single planting pattern over five years \cite{zhou_rhizospheric_2022} \\
    \hline
    172 & U,W & Microbial rewilding in the gut microbiomes of captive ring-tailed lemurs (Lemur catta) in Madagascar \cite{bornbusch_microbial_2022} \\
    \hline
    81 & U & Fecal level of butyric acid, a microbiome-derived metabolite, is increased in patients with severe carotid atherosclerosis \cite{sto_fecal_2022} \\
    \hline
    247 & U & Gut microbiota as an antioxidant system in centenarians associated with high antioxidant activities of gut-resident Lactobacillus \cite{wu_gut_2022} \\
    \hline
    20 & U & Helicobacter hepaticus augmentation triggers Dopaminergic degeneration and motor disorders in mice with Parkinson’s disease \cite{ahn_helicobacter_2023} \\
    \hline
    84 & W & Plant microbiomes harbor potential to promote nutrient turnover in impoverished substrates of a Brazilian biodiversity hotspot \cite{camargo_plant_2023} \\
    \hline
    17 & U,W & Increased levels of oral Streptococcus-derived d-alanine in patients with chronic kidney disease and diabetes mellitus \cite{nakade_increased_2022} \\
    \hline
    49 & W & Dynamics of rumen microbiome in sika deer (Cervus nippon yakushimae) from unique subtropical ecosystem in Yakushima Island, Japan \cite{eto_dynamics_2022} \\
    \hline
    3211 & W & The gut microbiota and depressive symptoms across ethnic groups \cite{bosch_gut_2022} \\
    \hline
    \end{tabular}
    \caption{\textbf{Latest 20 papers mentioning UniFrac with source “nature” on Google Scholar (Jan 2023).} UniFrac types are weighted (W) and unweighted (U).}
    \label{tab:papers}
\end{table}

\end{document}